\documentclass[runningheads]{llncs}

\usepackage{graphicx}
\usepackage{multirow}
\usepackage{csquotes}
\usepackage{amsmath}
\usepackage{array}
\usepackage{tabularx}
\newcolumntype{C}{>{\centering\arraybackslash}X}
\usepackage{tabularx}
\usepackage{booktabs}
\usepackage{algorithm}
\usepackage{algorithmic}
\usepackage{subcaption}

\usepackage{hyperref}

\begin{document}

\title{DSBA: Dynamic Stealthy Backdoor Attack with Collaborative Optimization in Self-Supervised Learning}

\titlerunning{DSBA: Dynamic Stealthy Backdoor Attack}
\authorrunning{DSBA: Dynamic Stealthy Backdoor Attack}
\author{Jiayao Wang\inst{1}\thanks{Corresponding Author. Email: \texttt{wjiayao0203@yzu.edu.cn}} \and
Mohammad Maruf Hasan\inst{1} \and
Yiping Zhang\inst{1} \and
Xiaoying Lei\inst{1} \and
Jiale Zhang\inst{1} \and
Qilin Wu\inst{2} \and
Junwu Zhu\inst{1} \and
Dongfang Zhao\inst{3}}

\institute{
School of Information and Artificial Intelligence, Yangzhou University, Yangzhou, China \and
School of Computing and Artificial Intelligence, Chaohu University, Chaohu, China \and
Tacoma School of Engineering and Technology, University of Washington, Tacoma, USA
}
\maketitle

\begin{abstract}
Self-Supervised Learning (SSL) has emerged as a significant paradigm in representation learning thanks to its ability to learn without extensive labeled data, its strong generalization capabilities, and its potential for privacy preservation. However, recent research reveals that SSL models are also vulnerable to backdoor attacks. Existing backdoor attack methods in the SSL context commonly suffer from issues such as high detectability of triggers, feature entanglement, and pronounced out-of-distribution properties in poisoned samples, all of which compromises attack effectiveness and stealthiness. To that, we propose a Dynamic Stealthy Backdoor Attack (DSBA) backed by a new technique we term Collaborative Optimization. This method decouples the attack process into two collaborative optimization layers: the outer-layer optimization trains a backdoor encoder responsible for global feature space remodeling, aiming to achieve precise backdoor implantation while preserving core functionality; meanwhile, the inner-layer optimization employs a dynamically optimized generator to adaptively produce optimally concealed triggers for individual samples, achieving coordinated concealment across feature space and visual space. We also introduce multiple loss functions to dynamically balance attack performance and stealthiness, in which we employ an adaptive weight scheduling mechanism to enhance training stability. Extensive experiments on various mainstream SSL algorithms and five public datasets demonstrate that: (i) DSBA significantly enhances Attack Success Rate (ASR) and stealthiness while maintaining downstream task accuracy; and (ii) DSBA exhibits superior robustness against existing mainstream defense methods.
\keywords{Self-Supervised Learning \and Backdoor Attack \and Collaborative Optimization \and Stealthiness }
\end{abstract}    
\section{Introduction}
\label{sec:intro}
Self-Supervised Learning (SSL)~\cite{SimCLR,MoCo} is a class of representation learning methods that automatically constructs supervisory signals from unlabeled large-scale data to exploit intrinsic data structures. By designing diverse pretext tasks for images or multimodal data, SSL acquires transferable and efficient feature representations during upstream pretraining, enhancing transferability and generalization performance in downstream tasks (e.g., classification, detection, and segmentation). By virtue of high data efficiency, low annotation costs and strong transferability, SSL has emerged as a research focus in unsupervised and transfer learning, achieving groundbreaking advances in Computer Vision (CV) and Natural Language Processing (NLP). However, the open nature and high generalization capability of SSL models also introduce novel security vulnerabilities.

In recent years, backdoor attacks have been proven as a prominent security threat capable of surreptitiously implanting malicious behaviors in supervised learning~\cite{I7}. Notably, this threat manifests more severely and co-mplexly in the SSL paradigm. Unlike traditional supervised scenarios that primarily poison downstream task models, SSL backdoor attacks primarily target upstream pretrained image encoders~\cite{GhostEncoder}. Attackers embed malicious backdoors into encoders either by directly training backdoored encoders or poisoning training data~\cite{CTRL,B6}. Such upstream backdoors exhibit exceptional stealthiness and cross-task transferability while maintaining normal model performance on clean inputs, yet trigger predetermined malicious outputs in downstream models when encountering specific triggers. Crucially, backdoor behaviors can be seamlessly inherited even when trigger samples are absent during downstream training, posing severe threats to SSL model security. Recent studies indicate that achieving highly stealthy, high-success-rate, and robust SSL backdoor attacks faces three core challenges:
\begin{itemize}
\item Image augmentations (e.g., RandomGrayscale and ColorJitter) widely adopted in SSL training substantially alter input distributions, causing feature-space overlap between backdoor samples and augmented samples ~\cite{I15}. This allows attack signals to be obscured by SSL objectives, resulting in unstable attack efficacy or failed implantation.
\item Existing methods struggle to balance stealth and high Attack Success Rate (ASR). Static/visible triggers (e.g., BadEncoder’s~\cite{BadEncoder} conspicuous patches) achieve high ASR (e.g., $99\%$ for BadEncoder) but are easily detected/reconstr-
ucted by manual inspection or automated defenses (e.g., STRIP~\cite{STRIP}, Neural Cleanse~\cite{NC}). Conversely, poisoning-based dynamic triggers (e.g., CTRL~\cite{CTRL}, BLTO~\cite{B6}.) offer improved stealth at the cost of suboptimal ASR (e.g., $61.90\%$ for CTRL under BYOL, $84.63\%$ for BLTO under SimSiam~\cite{SimSiam}).
\item Backdoor samples may exhibit detectable distributional deviations in embedding space~\cite{D18}, making them vulnerable to defenses based on distribution alignment (e.g., ASSET~\cite{ASSET}) or feature sanitization (e.g., Beatrix~\cite{Beatrix}).
\end{itemize}

To address the aforementioned challenges, although prior works such as MetaPoison~\cite{metapoison} introduce bi-level optimization for data poisoning and BLTO~\cite{B6} adapt it to SSL trigger generation, they are limited to supervised settings or trigger-only optimization, failing to jointly achieve global feature alignment and multi-dimensional stealthiness. This paper proposes a collaborative optimization-driven dynamic stealthy backdoor attack framework.~By synergizing outer-layer global optimization with inner-layer per-sample optimization, our method achieves sample-adaptive trigger generation and global distribution alignment, systematically enhancing attack effectiveness, stealthiness, and robustness. Specifically, the main contributions of this work are as follows:
\begin{itemize}
\item We systematically analyze and validate three critical challenges in SSL backdoor attacks: high trigger detectability, feature entanglement, and Out-of-Distribution Properties of backdoor samples, revealing their profound impact on attack stealthiness and effectiveness.
\item We propose the DSBA framework based on collaborative optimization, which is, to the best of our knowledge, the very first approach to decouple and co-optimize global feature space reshaping and per-sample dynamic trigger generation, thereby achieving unprecedented multi-dimensional stealthiness while maintaining high attack success rates.
\item We design multiple loss functions and adaptive weight scheduling mechanism, effectively integrating contrastive learning principles to dynamically balance attack efficacy with visual, feature-level, and distributional stealthiness under the collaborative framework, significantly enhancing training stability and attack robustness.
\item We conduct extensive experiments to compare the proposed DSBA method with state-of-the-art baselines across diverse self-supervised models and downstream tasks.
\end{itemize}

\section{Related Work}
\label{sec:Related Work}
\subsection{Self-Supervised Learning}
Self-Supervised Learning (SSL) trains universal feature extractors using large-scale unlabeled data, enabling efficient downstream task adaptation with minimal annotations. Mainstream SSL methods center on contrastive learning through three core strategies:
(1) Explicit contrastive strategies: Optimize feature distributions via positive/negative sample pairs, exemplified by SimCLR~\cite{SimCLR} (using InfoNCE loss) and MoCo~\cite{MoCo}
(employing momentum queues for negative sample management). (2) Implicit alignment mechanisms: Enhance cross-view consistency through dual-branch self-distillation, as in BYOL~\cite{BYOL} (predictor-head and momentum encoder co-training) and SimSiam (Stop-Gradient simplified architecture), both eliminating explicit negatives. (3) Online clustering optimization: Integrate clustering with feature learning to dynamically refine semantic boundaries end-to-end, typified by SwAV \cite{SwAV} (learnable cluster centers with Sinkhorn-Knopp soft assignment). These approaches significantly reduce annotation dependency while enhancing model adaptability to data distributions, establishing new paradigms for efficient training in vision, speech, and multimodal domains~\cite{D22}.

\subsection{Backdoor Attacks}
Traditional supervised backdoor attacks~\cite{BadNets} inject triggers to alter label associations. In SSL, however, attacks primarily target pre-trained encoders~\cite{PoisonedEncoder} by leveraging representation invariance: triggers entangle with target features under augmentations~\cite{BadEncoder}.
Based on trigger design and concealment, SSL backdoor attacks are categorized into: (1) Explicit triggers: e.g., BadEncoder~\cite{BadEncoder} employs visible patch triggers for data poisoning to precisely manipulate encoder responses. (2) Implicit triggers: e.g., CTRL~\cite{CTRL} employs augmentation-resistant triggers for feature entanglement; WaNet~\cite{WaNets} uses imperceptible image warping. GhostEncoder~\cite{GhostEncoder} utilizes steganographic embedding to generate backdoor samples from benign images, while IMPERATIVE~\cite{D35} decouples backdoor and augmented sample distributions while constraining trigger perceptibility. This paper proposes a collaborative optimization-driven dynamic stealthy backdoor attack method, aiming to achieve dynamic, stealthy, and highly robust backdoor attacks in SSL.

\subsection{Backdoor Defenses}
Backdoor attacks have emerged as a critical research focus in model security due to their high stealth and destructive potential, demanding urgent effective training-phase defenses. Existing countermeasures primarily encompass two categories: reverse-engineering-based approaches that extract trigger patterns and correlate them with malicious outputs for backdoor detection (e.g., Neural Cleanse~\cite{NC} and DECREE~\cite{D36}), though typically incurring substantial computational overhead, and sample-level detection methods that identify anomalous behaviors by analyzing input influence on model predictions—exemplified by STRIP~\cite{STRIP} detecting malicious samples via perturbation consistency, GradCAM~\cite{Grad-CAM} localizing trigger regions through activation maps. This paper employs these advanced defenses to systematically evaluate the proposed attack.
\section{Methodology}
\label{sec:Methodology}
This section focuses on introducing the threat model, observations and intuitions, and designing the collaborative backdoor attack framework while formulating its corresponding optimization objectives.

\subsection{Threat Model}
We characterize our threat model based on the attacker's objectives, background knowledge, and capabilities.\\
\textbf{Adversarial Objectives.} The attacker aims to construct a backdoored image encoder that implements covert malicious behavior while maintaining normal functionality, satisfying three key objectives:
(1) Attack Effectiveness Objective: Refers to dynamically generating triggers that precisely map to the target feature space to achieve high ASR.
(2) Multidimensional Stealthiness Objective: Requires imperceptibility across visual, statistical, and frequency domains.
(3) Transformation Robustness Objective: Ensures the attack remains effective under common data augmentations (e.g., cropping, color adjustments).\\
\textbf{Adversarial Knowledge and Capabilities.} The adversary, acting as a pretraining model provider, possesses full architectural control, and directs the pretraining pipeline—enabling training of conditional generative networks, designing collaborative optimization schemes, and configuring reconstruction loss functions with multidimensional stealth constraints and adaptive weight adjustment capabilities. However, lacking knowledge of downstream tasks or deployment interventions, the attacker can only implant backdoors during the single-stage pretraining phase while ensuring functional equivalence between backdoored and original encoders under constrained computational resources. We note that our attack is only applicable when downstream customers use an image encoder from an untrusted source.

\subsection{Observations and Intuitions}
\begin{figure*}[h]
  \centering
  \includegraphics[width=0.85\linewidth]{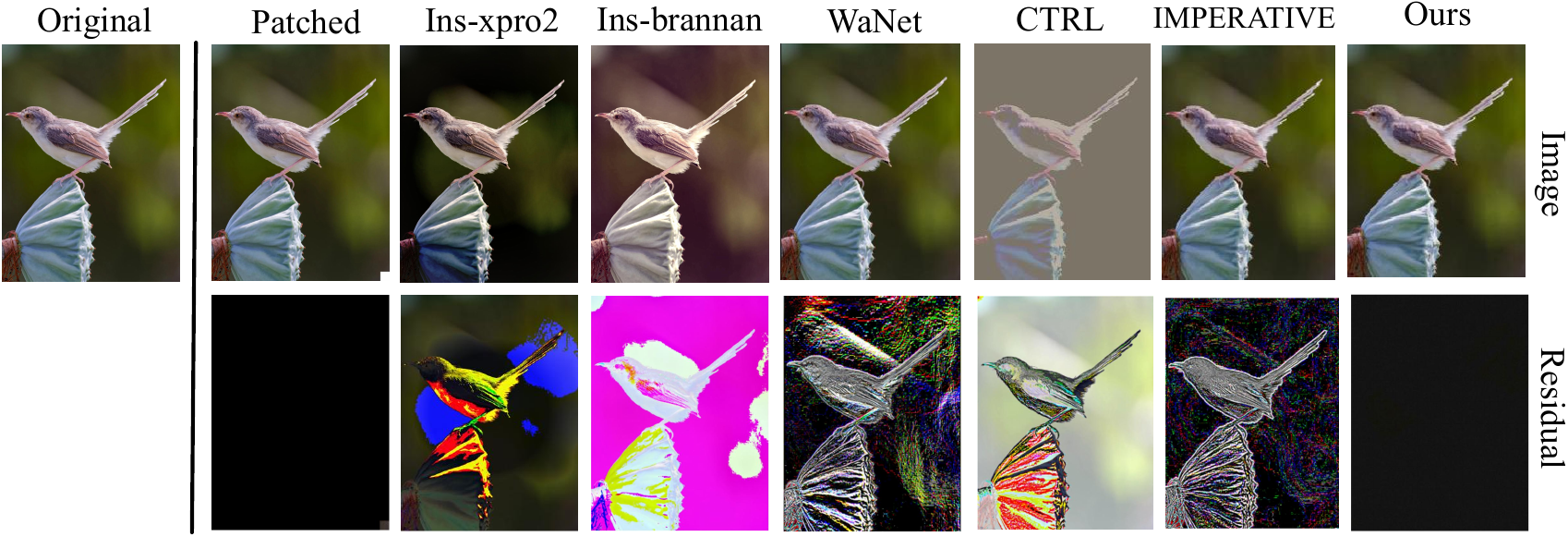}
   \caption{Comparison of clean, backdoored samples created by Patch trigger, Instagram filter trigger~\cite{Ins2} ~\cite{Ins1}, WaNet trigger~\cite{WaNets}, CTRL trigger~\cite{CTRL}, IMPERATIVE trigger~\cite{D35} and ours. Residuals are the difference between clean and backdoored images.}
   \label{fig:one}
\end{figure*}

\textbf{Observation I: Trigger Detectability.} Common SSL backdoor triggers exhibit poor concealment and are readily detectable. Specifically, these predominantly visible or static triggers are vulnerable to both manual and automated detection~\cite{GhostEncoder}. Sample-agnostic triggers can be easily reconstructed or identified by defenders via shared features~\cite{STRIP,NC}, making them susceptible to existing defenses. As shown in Figure~\ref{fig:one}, residual maps of current methods (e.g., Ins-xpro2~\cite{Ins1}, Ins-brannan~\cite{Ins2}) reveal visually evident artifacts, confirming the weak stealth, and high detectability of existing triggers.

\vspace{-10pt}
\vspace{-12pt}

\begin{figure}[htbp]
  \centering
  \begin{minipage}[b]{0.48\columnwidth}
    \centering
    \includegraphics[width=\textwidth]{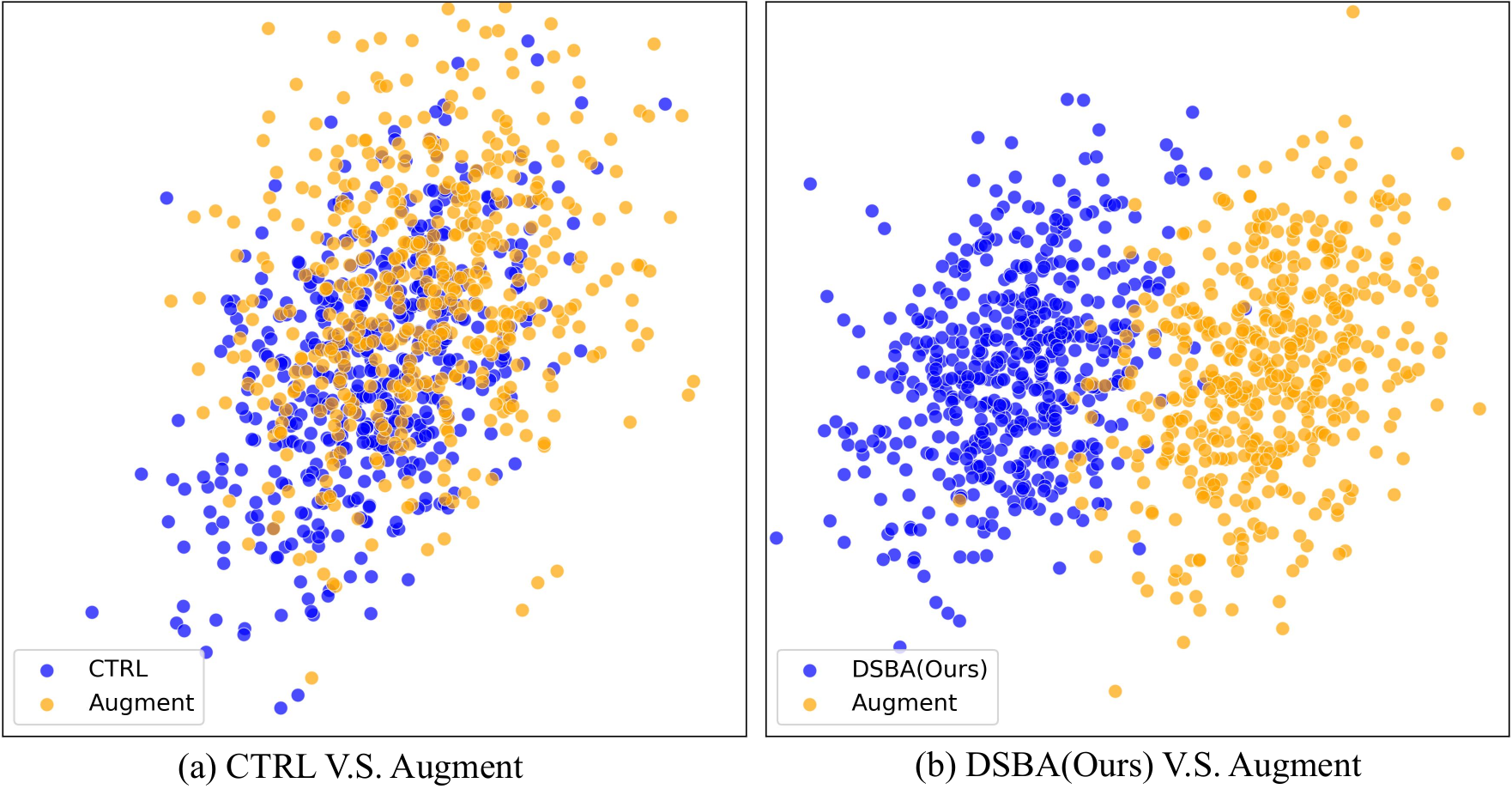}
    \caption{The t-SNE visualization of feature vectors in the latent space under different attacks.}
    \label{fig:two}
  \end{minipage}
  \hfill
  \begin{minipage}[b]{0.48\columnwidth}
    \centering
    \includegraphics[width=\textwidth]{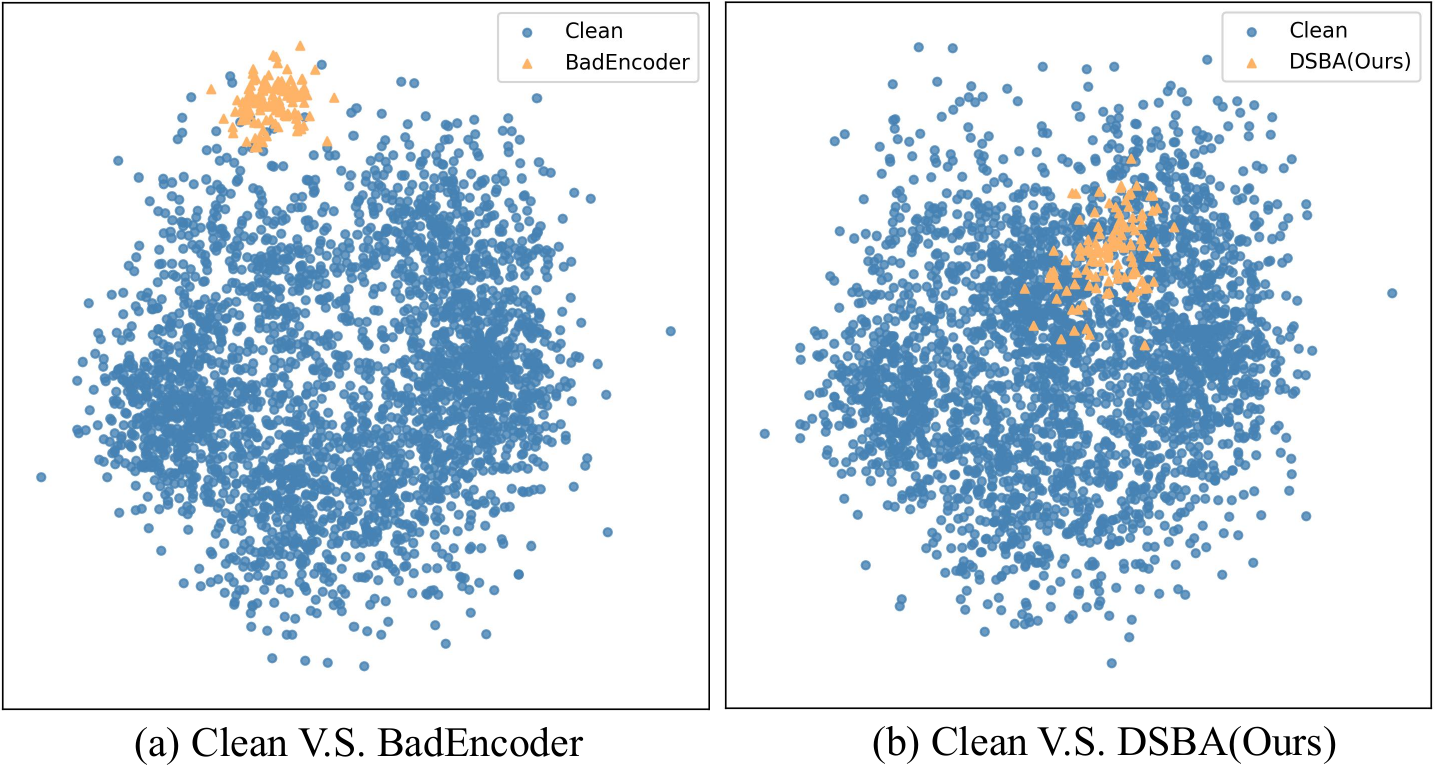}
    \caption{PCA visualization of clean and poisoned sample embeddings in backdoored models under different attacks.}
    \label{fig:three}
  \end{minipage}
\end{figure}

\noindent \textbf{Observation II: Feature Entanglement.} Augmented and backdoor samples exhibit highly entangled feature distributions. Specifically, a binary classification task using a pre-trained ResNet-18 model~\cite{resnet18} demonstrates the indistinguish ability between augmented samples and backdoor samples from the contrastive learning based covert attack CTRL. Both experimental results and t-SNE~\cite{t-SNE} visualizations (Figure~\ref{fig:two}(a)) confirm this feature entanglement. We infer that trigger-augmentation coupling would disrupt backdoor features, compromising the attack.\\
\textbf{Observation III: Out-of-Distribution Properties.} Existing backdoor samples exhibit out-of-distribution (OOD) properties. Tao et al.~\cite{D18} indicate that current SSL attack methods (e.g., backdoor attacks in contrastive learning) introduce pronounced backdoor signals into the embedding space, causing malicious samples' feature embeddings to deviate significantly from the normal data distribution. As illustrated by BadEncoder~\cite{BadEncoder}, PCA-reduced features~\cite{PCA} (Figure~\ref{fig:three}(a)) show poisoned samples exhibiting significant deviation from the benign distribution.

\vspace{-10pt}
\vspace{-12pt}

\noindent \textbf{Intuition and Design Motivation.} Based on the above observations, issues such as trigger detectability, feature entanglement, and out-of-distribution properties significantly constrain the effectiveness and stealthiness of backdoor attacks. To systematically address these challenges, we propose a collaborative dynamic stealthy backdoor attack framework driven by outer-inner co-optimization. Specifically:\\
\textbf{To address trigger detectability}, the inner layer dynamically adjusts the trigger's visual imperceptibility through $L_{ste}$ and $L_{cons}$ constraints, while the outer layer introduces $L_{perc}$ to enhance visual consistency of backdoor samples and reduce detection risks.\\
\textbf{For feature entanglement}, the inner layer employs $L_{eff}$ to ensure adaptive trigger generation per sample, improving attack precision and robustness, with the outer layer further enhancing attack capability via $L_{align}$ while preserving model functionality.\\
\textbf{Regarding out-of-distribution properties}, the outer layer designs $L_{dist}$ combining Jensen-Shannon divergence and statistical moment matching, aligning backdoor sample distribution with clean samples to enhance stealthiness and anti-detection capability.

\subsection{Collaborative Attack Framework Design}
\begin{figure}[h]
  \centering
  \includegraphics[width=0.65\linewidth]{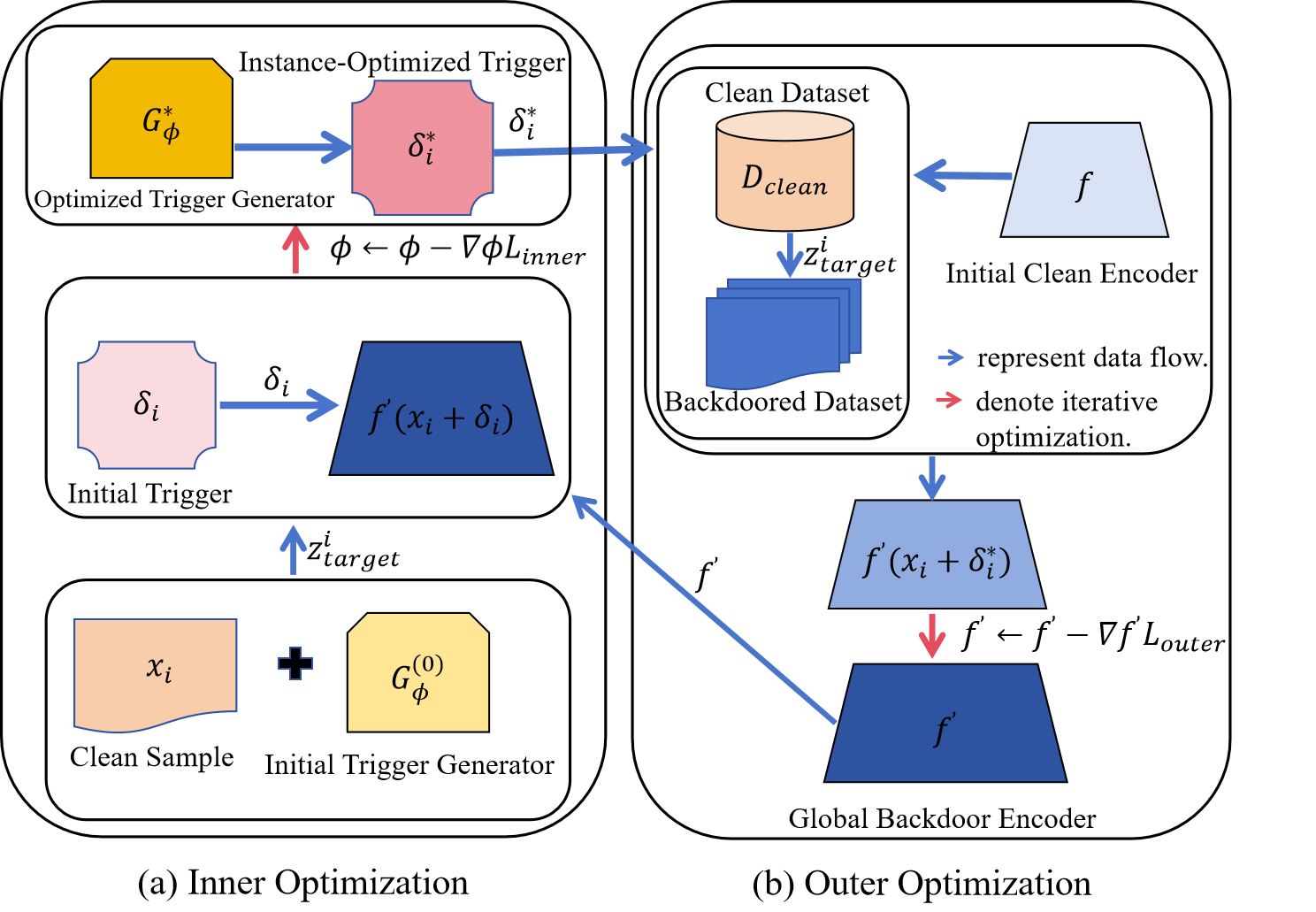}
   \caption{Inner-Outer layer collaborative optimization for DSBA. The Inner Optimization updates the trigger generator parameters $\phi$ by minimizing $L_{inner}$, producing sample-specific optimal triggers $\delta_i^*$ ($\delta_i$ evolves into $\delta_i^*$). The Outer Optimization uses these triggers to update the backdoor encoder $f'$ by minimizing $L_{outer}$. The two layers form a closed loop, alternating updates for synergistic attack and stealth.}
   \label{fig:four}
\end{figure}

\textbf{Collaborative optimization framework overview.} 
Figure~\ref{fig:four} depicts the overall framework of our dynamic stealthy backdoor attack driven by inner-outer collaborative optimization. This framework partitions the attack process into two components: the outer layer (global backdoor encoder optimization) and the inner layer (per-sample dynamic trigger generation), which are responsible for global attack capability and individual stealthiness respectively. The outer-layer optimization focuses on holistic attack effectiveness and distributional stealth, while the inner-layer optimization addresses per-sample attack precision and robustness. Through collaborative optimization across both layers, we achieve efficient and covert backdoor attacks. (The collaborative inner-outer optimization mechanism and the pseudocode for this algorithm are detailed in Supplementary Sec. 2 and 6, respectively.)


\vspace{-18pt}
\vspace{-19pt}

\noindent \textbf{Outer optimization: Global backdoor encoder optimization.} The outer-layer optimization adjusts parameters of the backdoor encoder, whose core consists of a clean encoder $f$, a backdoor encoder $f'$, and a dynamic trigger generator $G_\phi$. Utilizing the optimal dynamic triggers $\delta_i^* = G_\phi(x_i)$ generated by inner-layer optimization as input, it embeds triggers into original samples $x_i$ to produce backdoored samples $x_i+\delta_i^*$. Three pivotal loss functions are designed to synergistically achieve global attack effectiveness, visual stealth, and functionality preservation. The design and roles of each outer-layer loss function will be detailed subsequently.\\
\textbf{Dual alignment loss.} Given that SSL models generate feature representations rather than relying on explicit labels, the attacker ensures that backdoored samples successfully mimic the target semantics to achieve the attack objective, while maintaining normal model performance on clean inputs to evade detection, thereby enabling efficient and stealthy backdoor attacks. We term this process dual feature alignment, referring to the alignment of backdoored samples with target-class features, and the feature alignment of clean samples between the backdoored encoder and clean encoder. Following BadEncoder~\cite{BadEncoder}, the attacker's dual feature alignment can be formally expressed as:
\begin{small}
\begin{align}
L_{align} = -\frac{1}{|D_s|} \sum_{x_i \in D_s} s\left(f'(x_i + \delta_i^*), z_{target}^i\right) - \lambda \frac{1}{|D_{clean}|} \sum_{x \in D_{clean}} s\left(f'(x), f(x)\right),
\end{align}
\end{small}%
where $s(\cdot,\cdot)$ denotes the cosine similarity function, $\lambda$ balances the two objectives, $D_s$ represents the shadow dataset, and $D_{clean}$ is the clean sample set. For each target class, we compute its reference feature $z_{target}^i$ by aggregating the features of multiple reference inputs(The computation procedure is detailed in Supplementary Sec. 1.1.). \\
\textbf{Perceptual loss.} To enhance visual stealth of backdoored samples, a multi-layer perceptual loss is designed that computes weighted cosine similarity using normalized features extracted from various layers of networks (e.g., VGG), incorporating multi-scale feature constraints to ensure high perceptual consistency between backdoored and original samples. The loss function is defined as:
\begin{small}
\begin{equation}
  L_{perc} = \frac{1}{|D_s|} \sum_{x_i \in D_s} \sum_{l} \alpha_l \left[1 - \frac{\phi_l(x_i)}{\|\phi_l(x_i)\|_2} \cdot \frac{\phi_l(x_i + \delta_i^*)}{\|\phi_l(x_i + \delta_i^*)\|_2}\right],  
\end{equation}
\end{small}%
the multi-scale perceptual loss enhancement can be expressed as:
\begin{small}
\begin{equation}
L^{multi-scale}_{perc} = \sum_{S \in \{1, 0.5, 0.25\}} \beta_s \cdot L^{(s)}_{perc} ,
\end{equation}
\end{small}%
where $\phi_l(x_i)$ denotes the feature extraction from the $l$-th layer of the VGG network, and $\alpha_l$ represents the weighting coefficient for different layers. The multi-scale perceptual loss is achieved through weighted summation of losses at varying scaling ratios. $S$ denotes the set of image scaling factors (e.g., original, 0.5×, 0.25×) for multi-scale perceptual consistency, while $\beta_s$ indicates the weighting coefficient for losses at different scales. 
$L^{(s)}_{perc}$ denotes the perceptual loss computed at scale $s \in S$.\\
\textbf{Distribution alignment loss.} To address the issue where backdoored samples manifest as out-of-distribution anomalies in feature space and are easily detected, we design loss $L_{dist}$ that constrains feature distributions between backdoored and clean samples using Jensen-Shannon divergence and statistical moment matching (mean, variance). This prevents backdoored samples from exhibiting outlier characteristics in feature space. The loss function is defined as follows:
\begin{small}
\begin{equation}
\begin{aligned}
L_{dist} =\ & \lambda_{js} \cdot D_{JS}(P_{clean} \parallel P_{poison}) + \lambda_{stat} \cdot \\
&  \left[ \|\mu_{clean} - \mu_{poison}\|_2^2 + \|\sigma_{clean} - \sigma_{poison}\|_2^2 \right],
\end{aligned}
\end{equation}
\end{small}%
where $D_{JS}$ denotes the Jensen-Shannon divergence measuring the overall similarity between two distributions, 
$\lambda_{js}$ and $\lambda_{stat}$ are weighting coefficients controlling the contributions of the divergence term and statistical moment matching term to the total loss, respectively. The variables $P_{clean}$ and $P_{poison}$
 represent the feature distributions of clean and backdoored samples, while 
$\mu_{clean}$ and $\mu_{poison}$ indicate their distribution means,  $\sigma_{clean}$ and $\sigma_{poison}$ their distribution variances. The operator $\|\cdot\|_2^2$ computes the squared $L2$ norm for measuring distances between means and variances.

Based on the aforementioned loss formulations and design objectives, we formulate the holistic outer-layer backdoor attack as an optimization problem. Specifically, our dynamic stealthy backdoor encoder constitutes the solution to the following optimization problem:
\begin{equation}
  \min_{f'}{L_{outer}} = \omega_1 \cdot L_{align} + \omega_2 \cdot L_{perc} + \omega_3 \cdot L_{dist},
\end{equation}%
where $\omega_i$ denotes the weight vector for individual losses, dynamically adjustable via an adaptive scheduling mechanism. (Details of the adaptive scheduling mechanism are provided in Supplementary Sec. 1.2.) \\
\textbf{Inner optimization: Dynamic stealthy trigger generator optimization.} The objective of inner-layer optimization is to generate an optimal dynamic trigger for each sample, ensuring precise alignment of backdoored sample features with target features while maintaining visual stealth and robustness against data augmentations. To this end, three key loss functions are designed. By minimizing the overall inner-layer loss function, we ultimately obtain a Generator $G_\phi$ capable of producing high-quality dynamic stealthy triggers for arbitrary samples. We subsequently detail the design and roles of each loss function in the inner-layer optimization.\\
\textbf{Sample-level effectiveness loss.} To ensure precise feature alignment of each backdoored sample with the target feature space for effective individual attacks while enhancing robustness against content variations, we design a sample-level efficacy loss. This loss comprises two components: (1) minimizing the cosine distance between backdoored sample features and target features; and (2) constraining the robustness of generated triggers to sample content perturbations. The loss function is defined as follows:

\begin{equation}
\begin{aligned}
L_{eff} =
    \ & -\Bigg[ ( 1 - S_{target} )+\lambda_{temp} \exp\left(-\frac{1 - S_{target}}{\tau}\right) \Bigg] \\
&- \lambda_{content} \left\|G_\phi(x_i) - G_\phi(\mathrm{Aug}(x_i))\right\|_2^2,
\end{aligned}
\end{equation}
where $S_{target}=\frac{f'(x_i + \delta_i)}{\|f'(x_i + \delta_i)\|_2} \cdot \frac{z^i_{target}}{\|z^i_{target}\|_2}$ denotes the cosine similarity between the backdoor sample feature and the target feature, $\mathrm{Aug}(x_i)$ denotes light data augmentation applied to $x_i$, $\lambda_{temp}$ and $\lambda_{content}$
represent the weighting coefficients for the temperature adjustment term and content robustness term respectively, and $\tau$ is the temperature coefficient, the variable $\delta_i$ denotes the trigger to be further optimized.\\
\textbf{Visual stealth loss.} To enhance the visual imperceptibility of backdoored samples and ensure high visual consistency with original samples, we design a visual stealth loss. This loss integrates Structural Similarity (SSIM)~\cite{D46} and L2 regularization to constrain visual fidelity between backdoored and original samples, thereby mitigating detection risks. The loss function is defined as follows:
\begin{equation}
L_{ste} = \alpha \cdot (1 - \text{SSIM}(x_i, x_i + \delta_i)) + \beta \cdot \|\delta_i\|_2^2,
\end{equation}%
where $\alpha$ and $\beta$ denote the weighting coefficients for the structural similarity term and magnitude regularization term respectively, enabling optimal balancing of both loss components. SSIM represents the structural similarity constraint.\\
\textbf{Constraint loss.} To enhance the adaptability and stealth of trigger patterns against common data transformations, we design a constraint preservation loss. This loss restricts trigger amplitude and structure through $L_\infty$ perturbation bounds, spatial smoothness, and frequency-domain sparsity, thereby improving robustness and imperceptibility. The loss function is formulated as follows:
\begin{equation}
\begin{aligned}
L_{cons} =\ & \lambda_{norm} \cdot \max(0, \|\delta_i\|_\infty - \epsilon) \\
& +\ \lambda_{sm} \cdot TV(\delta_i) + \lambda_{freq} \cdot \|FFT(\delta_i)\|_1,
\end{aligned}
\end{equation}%
where $\lambda_{norm}$, $\lambda_{sm}$ and $\lambda_{freq}$ denote the weighting coefficients for the $L_\infty$ constraint, spatial smoothness, and frequency-domain sparsity terms, respectively. The $L_\infty$ constraint function ensures $\max(\cdot)$
 pixel-wise perturbations do not exceed budget $\epsilon$ (typically 8/255), imposing a penalty when perturbations exceed the budget and zero otherwise. $FFT(\cdot)$ represents the Fourier transform and $TV(\cdot)$ denotes total variation regularization.
 
 Building on the aforementioned loss formulations and design objectives, we obtain a generator $G_\phi$ capable of adaptively generating dynamic covert triggers for each sample. The composite loss function for the inner-layer optimization is defined as:

\begin{equation}
\min_{G_\phi} L_{inner} = \sum_i \left[ \mu_1 \cdot L_{eff} + \mu_2 \cdot L_{ste} + \mu_3 \cdot L_{cons} \right],
\end{equation}%
where $\mu_i$ denotes the respective weighting coefficients for each loss component.
\section{Evaluation}
To demonstrate the effectiveness and stealthiness of our method, we implemented DSBA using PyTorch and compared its performance with existing state-of-the-art backdoor attack methods. All experiments were conducted on NVIDIA A100 GPUs with 80GB of memory, and each experiment was repeated five times under independent runs, with their average results reported. We designed comprehensive experiments to address the following three research questions:\\
\textbf{RQ1 (Effectiveness of DSBA)}: Can DSBA successfully inject dynamic stealthy backdoors into SSL?\\
\textbf{RQ2 (Stealthiness of DSBA)}: Can DSBA achieve good
stealthiness and naturalness under different evaluation metrics?\\
\textbf{RQ3 (Robustness of DSBA)}: Can DSBA effectively resist
existing defense methods?
\subsection{Experimental Setup}
\label{sec:Experimental Setup}
\textbf{Datasets.} Experiments utilize five datasets:~CIFAR-10~\cite{D41}, STL-10~\cite{D42}, GTSRB~\cite{D43}, SVHN~\cite{D44}, and TinyImageNet~\cite{D45}. By default, we employ CIFAR-10 for upstream pre-training and STL-10 for downstream evaluation. See Supplementary Sec. 3 for dataset details.\\
\textbf{Evaluation Metrics.} Following BadEncoder~\cite{BadEncoder} and DRUPE~\cite{D18}, Attack Effectiveness: Assessed by Clean Accuracy (CA), Attack Success Rate (ASR), and Backdoor Accuracy (BA) to evaluate performance on normal tasks, triggered attacks, and overall functionality, respectively. Stealthiness and Naturalness: Evaluated across structural similarity, noise level, perceptual difference, feature similarity, and distribution consistency using SSIM~\cite{D46}, PSNR~\cite{D47}, LPIPS~\cite{D48}, FSIM~\cite{D49}, and FID~\cite{D50}.
 where higher SSIM/PSNR/FSIM and lower LPIPS/FID indicate superior stealth. (See Supplementary Sec. 4 for details.)\\
\textbf{Baselines.} DSBA is compared against SOTA methods: IMPERATIVE~\cite{D35}, GhostEncoder~\cite{GhostEncoder}, WaNet~\cite{WaNets}, and CTRL~\cite{CTRL}. All baselines are adapted to the SSL scenario per experimental protocols in BadEncoder~\cite{BadEncoder} and strictly follow original implementations.\\
\textbf{Implementation Details.} We adopt SimCLR for SSL, with ResNet-18 as the encoder and a two-layer MLP as the predictor. Key parameters include a momentum coefficient $m$ = 0.99, batch size $B$ = 256, and SGD optimizer with learning rate $lr$ = 0.001. During the upstream backdoor injection phase, a U-Net~\cite{D51} serves as the trigger generator and a PatchGAN~\cite{D52} as the discriminator, both optimized using Adam. Training proceeds for 200 epochs: The initial phase fixes the encoder while optimizing the generator, followed by an alternating optimization scheme where the generator updates every 3 epochs and the encoder updates every epoch. Our approach incurs twice the training cost of BadEncoder~\cite{BadEncoder} ($\sim$10 hours, our pre-training uses a single A100 GPU; with greater computational resources, attackers could substantially reduce pre-training time.) during upstream training, with zero cost during inference after deployment.

\vspace{-10pt}
\begin{table*}[htbp]
\centering
\caption{Camparison of attack performance on different datasets. The best result are  \textbf{highlighted}.}
\label{tab:one}
\scalebox{0.75}{
\begin{tabular}{l|c|ccccccccccc@{}}
\hline
\textbf{Pre-training} & \textbf{Downstream} & \textbf{Benign} & \multicolumn{2}{c}{\textbf{WaNet}} & \multicolumn{2}{c}{\textbf{CTRL}} & \multicolumn{2}{c}{\textbf{GhostEncoder}} & \multicolumn{2}{c}{\textbf{IMPERATIVE}} & \multicolumn{2}{c}{\textbf{Ours}} \\
\cline{3-13}
\textbf{Dataset} & \textbf{Dataset} & CA$\uparrow$ & BA$\uparrow$ & ASR$\uparrow$ & BA$\uparrow$ & ASR$\uparrow$ & BA$\uparrow$ & ASR$\uparrow$ & BA$\uparrow$ & ASR$\uparrow$ & BA$\uparrow$ & ASR$\uparrow$ \\
\hline
         & STL-10 & 77.42 & 72.17 & 12.17 & 74.46 & 65.56 & 75.28 & 80.17 & 73.25 & 87.16 & 76.52 & \textbf{96.76} \\
CIFAR-10 & GTSRB & 80.14 & 73.24 & 8.25 & 75.43 & 63.27 & 78.32 & 65.14 & 72.18 & 82.45 & 78.23 & \textbf{93.14} \\
         & SVHN & 65.43 & 55.75 & 13.15 & 58.24 & 46.18 & 65.14 & 35.78 & 56.42 & 75.13 & 64.74 & \textbf{94.85} \\
\hline
       & CIFAR-10 & 83.22 & 81.42 & 11.48 & 75.19 & 62.33 & 81.56 & 79.14 & 81.35 & 76.74 & 84.15 & \textbf{99.67} \\
STL-10 & GTSRB & 77.34 & 78.24 & 3.95 & 69.37 & 63.04 & 71.35 & 66.12 & 73.22 & 79.47 & 75.24 & \textbf{96.43} \\
       & SVHN & 57.46 & 54.56 & 16.43 & 55.78 & 50.17 & 52.48 & 34.83 & 60.13 & 80.12 & 60.16 & \textbf{95.74} \\
\hline
              & STL-10 & 85.53 & 82.57 & 12.75 & 80.14 & 47.18 & 81.34 & 51.43 & 77.63 & 79.14 & 84.15 & \textbf{98.14} \\
Tiny-ImageNet & GTSRB & 78.32 & 75.65 & 9.71 & 71.12 & 44.75 & 72.48 & 47.16 & 71.32 & 76.53 & 76.12 & \textbf{96.28} \\
             & SVHN & 74.65 & 71.43 & 14.26 & 61.95 & 41.28 & 68.42 & 39.47 & 60.48 & 71.43 & 72.18 & \textbf{93.14} \\
\hline
\end{tabular}
}
\end{table*}
\vspace{-17pt}

\subsection{Effectiveness Evaluation (RQ1)}
\begin{figure}[h]
  \centering
  \includegraphics[width=0.7\linewidth]{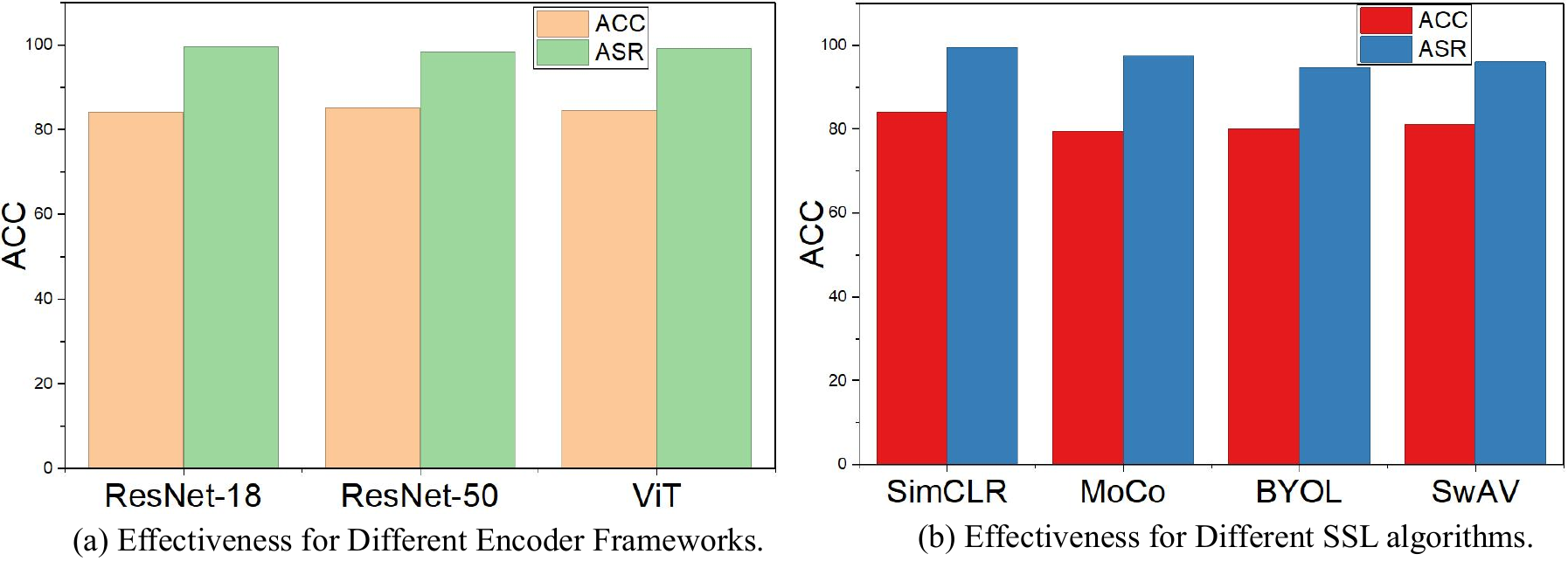}
   \caption{Experimental results for different encoder architectures and SSL algorithms.}
   \vspace{-5mm}
   \label{fig:five}
\end{figure}

\textbf{Effectiveness comparison with SOTA attack methods.} Table~\ref{tab:one} compares DSBA with four SOTA attack methods under standard SSL settings, demonstrating DSBA’s superior ASR and BA in all data sets. For instance, with STL-10 pretraining and CIFAR-10 downstream, DSBA achieves $99.67\%$ ASR and $84.15\%$ BA, significantly outperforming WaNet~\cite{WaNets}, IMPERATIVE~\cite{D35}, GhostEncoder~\cite{GhostEncoder}, and CTRL~\cite{CTRL}.\\
\textbf{Accuracy Preservation.} DSBA effectively decouples backdoor samples from augmented features (Figure~\ref{fig:two}(b)), ensuring high attack success while minimally impacting clean-sample classification. As shown in Table~\ref{tab:one}, BA and CA differ by less than $3\%$ across dataset pairs (e.g., $76.52\%$ BA vs. $77.42\%$ CA for CIFAR-10 pretraining → STL-10 downstream).\\
\textbf{Effectiveness on different encoder architectures.} The attack successfully implants backdoors while maintaining high BA across diverse encoders including ResNet-18~\cite{resnet18}, ResNet-50~\cite{resnet18}, and ViT~\cite{D54} (Figure~\ref{fig:five}(a)), further confirming its architecture-agnostic generalization.\\
\textbf{Effectiveness on different SSL algorithms.} DSBA exhibits consistent high performance across four mainstream SSL methods—SimCLR~\cite{SimCLR}, MoCo~\cite{MoCo}, BYOL~\cite{BYOL}, and SwAV~\cite{SwAV} (Figure~\ref{fig:five}(b)), validating its adaptability and generalization.

\vspace{-10pt}
\vspace{-15pt}

\vspace{-10pt}
\begin{table}[htbp]
\centering
\caption{Stealthness evaluation on different methods.}
\label{tab:two}
\scalebox{0.75}{
\begin{tabular}{@{}l|ccccc@{}}
\hline
\textbf{Methods} & \textbf{SSIM$\uparrow$} & \textbf{PSNR$\uparrow$} & \textbf{LPIPS$\downarrow$} & \textbf{FSIM$\uparrow$} & \textbf{FID$\downarrow$} \\
\hline
BadEncoder~\cite{BadEncoder} & 0.8217 & 14.1311 & 0.07429 & 0.837 & 54.165 \\
CTRL~\cite{CTRL} & 0.9125 & 31.4783 & \textbf{0.00057} & 0.876 & 71.145 \\
WaNet~\cite{WaNets} & 0.7761 & 15.2713 & 0.07514 & 0.654 & 97.996 \\
Ins-Kelvin~\cite{Ins-Kelvin} & 0.4877 & 17.1355 & 0.15022 & 0.633 & 96.233 \\
Ins-Xpro2~\cite{Ins1} & 0.5842 & 18.9145 & 0.04322 & 0.823 & 36.128 \\
SSLBKD~\cite{SSLBKD} & 0.8725 & 16.5314 & 0.09417 & 0.886 & 115.422 \\
IMPERATIVE~\cite{D35} & 0.9542 & 35.7435 & 0.00884 & 0.924 & 16.357 \\
Ours & \textbf{0.9918} & \textbf{36.8312} & 0.00421 & \textbf{0.973} &   \textbf{14.425} \\
\hline
\end{tabular}
}
\end{table} 
\vspace{-20pt}

\subsection{Stealthiness Evaluation (RQ2)}
\textbf{Stealthiness Evaluation Metrics.} Figure~\ref{fig:one} and Table~\ref{tab:two} demonstrate that DSBA-generated poisoned samples exhibit minimal residuals with negligible artifacts, outperforming SOTA methods across all five metrics (SSIM~\cite{D46} 0.9918, PSNR~\cite{D47} 36.8312, LPIPS~\cite{D48} 0.00421, FSIM~\cite{D49} 0.973, FID~\cite{D50} 14.425), indicating near-imperceptible perturbations. While CTRL~\cite{CTRL} shows marginally better LPIPS, DSBA achieves superior performance on critical metrics like SSIM and PSNR, coupled with substantially higher average ASR ($96.02\%$ vs. CTRL’s $53.75\%$), thus balancing high stealth and attack efficacy.\\
\textbf{Stealthiness from the Latent Space Perspective.} Many backdoor defense methods operate under the assumption that poisoned and benign samples are separable in the latent space. It is therefore crucial to ensure the stealthiness of an attack from this perspective. We visualize the feature embeddings of BadEncoder~\cite{BadEncoder} using PCA in Figure~\ref{fig:three}, where two distinct clusters are observed, indicating easy detectability by clustering algorithms. In contrast, with our DSBA, the features of poisoned samples blend seamlessly with those of benign ones, forming a single cluster. This demonstrates that DSBA achieves optimal stealth in the latent space, breaking the separability assumption and effectively evading backdoor defenses.~(Additional stealthiness evaluation in Supplementary Sec. 5).

\subsection{Robustness Evaluation (RQ3)}
To evaluate DSBA's resilience to mainstream defenses, we implemented and tested DECREE~\cite{D36}, Neural Cleanse (NC)~\cite{NC}, Beatrix~\cite{Beatrix}, STRIP~\cite{STRIP}, and GradCAM~\cite{Grad-CAM}.  \\
\textbf{Resistance to DECREE.} DECREE detects backdoored encoders via trigger inversion, flagging models with $\mathcal{PL}^1-Norm <0.1$. DSBA successfully evades detection as all generated triggers exhibit $\mathcal{PL}^1-Norm >0.1$ (Table~\ref{tab:three} (a)).\\
\textbf{Resistance to Neural Cleanse (NC).} Neural Cleanse identifies backdoors through anomaly indices ($>2$ threshold). Evaluated on downstream classifiers, DSBA maintains anomaly indices $<2$ across all datasets (Table \ref{tab:three} (a)), escaping detection.\\
\textbf{Resistance to Beatrix.} Beatrix detects poisoned samples via feature-space anomalies. While achieving $>96.9\%$ detection against BadEncoder~\cite{BadEncoder}, Beatrix shows $<50\%$ accuracy against DSBA on CIFAR-10/STL-10 (Table \ref{tab:three} (b))—appro-aching random guessing. Analysis reveals near-identical deviation distributions between DSBA-poisoned and clean samples.

\vspace{-10pt}
\begin{table*}[htbp]
\centering
\caption{Defense evaluation results.}
\label{tab:three}
\vspace{-8pt}
\begin{subtable}{0.48\textwidth}
\caption{Detection results by NC and DECREE}
\centering
\resizebox{!}{0.95cm}{
 \begin{tabular}{@{}l|c|cc@{}}
    \hline
    \textbf{Pre-training} & \textbf{Downstream} &  \textbf{NC}~\cite{NC} & \textbf{DECREE}~\cite{D36} \\
    \cline{3-4}
    \textbf{Dataset} & \textbf{Dataset} & Anomaly Index & $\mathcal{PL}^1$-Norm \\
    \hline
         & STL-10 & 1.05 & 0.24  \\
CIFAR-10 & GTSRB & 1.21 & 0.36  \\
         & SVHN & 1.36 & 0.15 \\
   \hline
       & CIFAR-10 & 0.88 & 0.24 \\
STL-10 & GTSRB & 1.12 & 0.32 \\
       & SVHN & 1.35 & 0.26 \\
   \hline
   \end{tabular}
}
\end{subtable}%
\begin{subtable}{0.42\textwidth}
\caption{Detection results by Beatrix}
\resizebox{!}{0.8cm}{
\centering
 \begin{tabular}{@{}c|c|ccccc@{}}
    \hline
    \textbf{Datasets} & \textbf{Methods} & \textbf{TP} & \textbf{FP} & \textbf{FN} & \textbf{TN} & \textbf{ACC (\%)} \\
    \hline
    \multirow{2}{*}{CIFAR-10} & BadEncoder & 497 & 22 & 6 & 499 & 97.27 \\
           & Ours & 0 & 22 & 500 & 480 & 47.90 \\
    \hline
    \multirow{2}{*}{STL-10} & BadEncoder & 493 & 22 & 6 & 480 & 97.20 \\
         & Ours & 4 & 22 & 493 & 480 & 48.45 \\
    \hline
    \end{tabular}
}
\end{subtable}

\end{table*}
\vspace{-13pt}
\vspace{-10pt}
\begin{figure}[htbp]
  \centering
  \begin{minipage}[b]{0.53\columnwidth}
    \centering
    \includegraphics[width=0.95\linewidth]{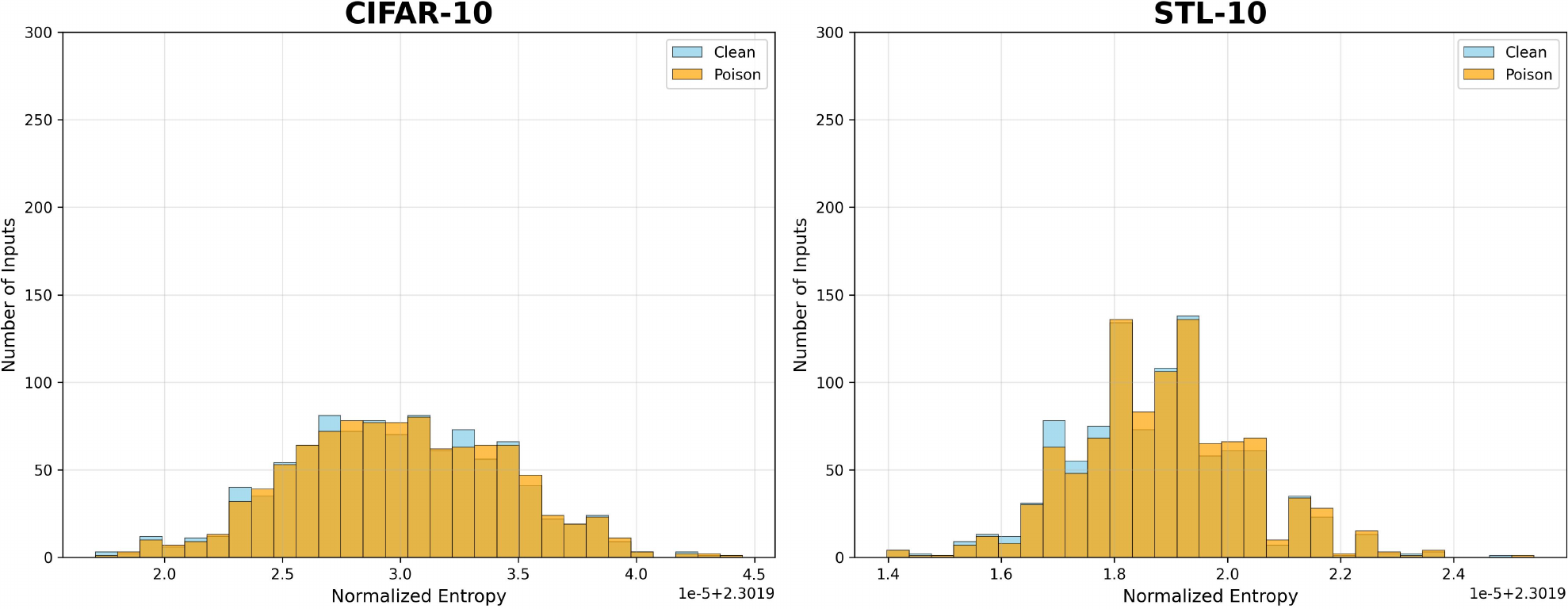}
    \caption{Experimental results of STRIP.}
    \label{fig:six}
  \end{minipage}
  \hfill
  \begin{minipage}[b]{0.43\columnwidth}
    \centering
    \includegraphics[width=0.9\linewidth]{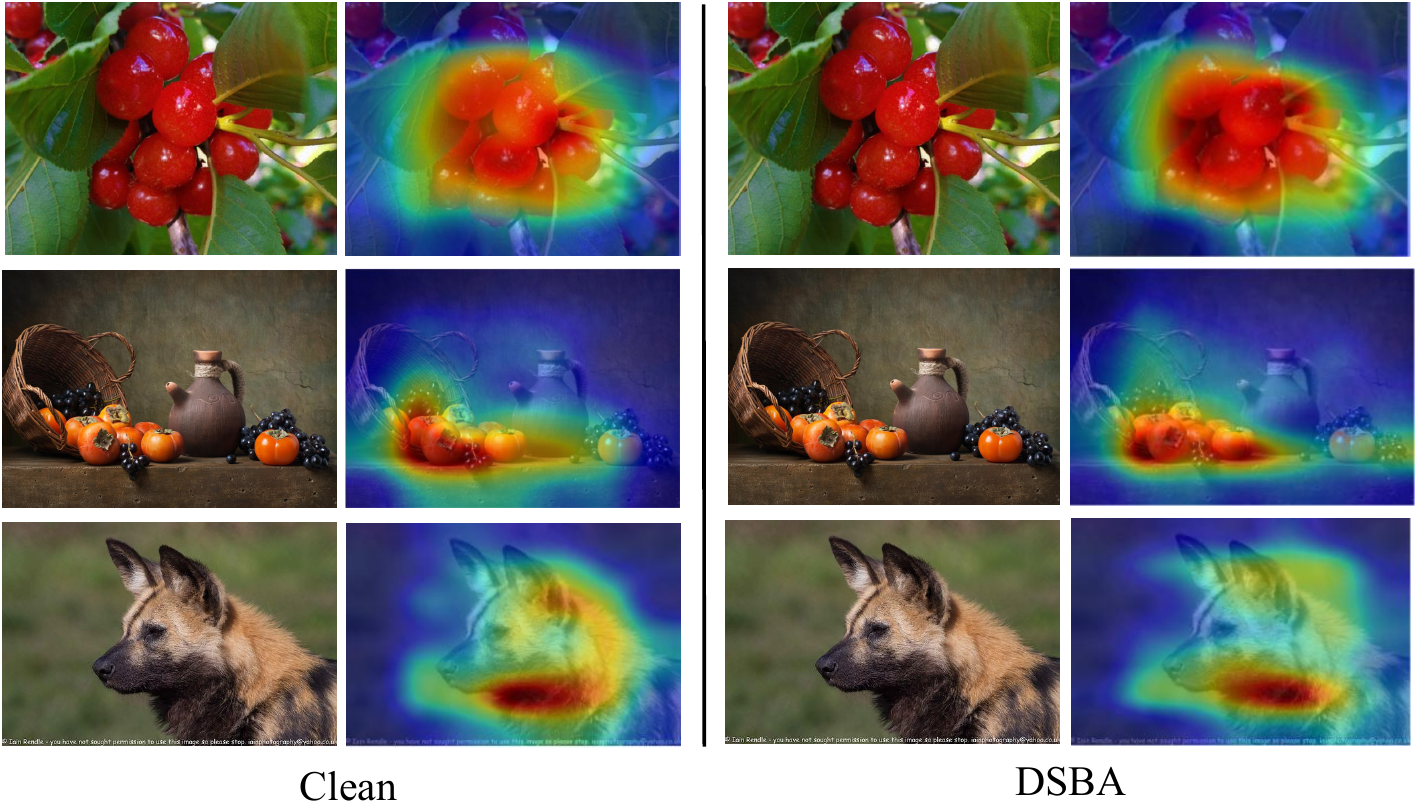}
    \caption{GradCAM visualization results for both clean and backdoored models.}
    \vspace{-3mm}
    \label{fig:seven}
  \end{minipage}
\end{figure}
\vspace{-13pt}

\noindent \textbf{Resistance to STRIP.} STRIP is a sample-based backdoor detection method that operates on the hypothesis that backdoored models exhibit prediction stability on malicious samples. It detects such samples by computing the entropy after superimposing random clean samples. As shown in Figure~\ref{fig:six}, the entropy distributions of clean and poisoned samples overlap almost completely, rendering STRIP ineffective at distinguishing between them. This result demonstrates that our dynamically covert backdoor trigger is highly concealed and successfully evades STRIP's detection mechanism.\\
\textbf{Resistance to GradCAM.} GradCAM generates heatmaps to visualize pixel-wise contributions to model predictions, with anomalous heatmaps often indicating backdoored models. As shown in Figure~\ref{fig:seven}, the visualized heatmaps of the clean model and the DSBA-compromised backdoored model exhibit high similarity. This demonstrates that DSBA effectively evades GradCAM-based detection by maintaining heatmap consistency with benign models.

\subsection{Ablation Study}
This section validates the critical roles of loss functions in the collaborative optimization framework through ablation studies, where \enquote{All losses} denotes enabling all six losses and \enquote{No $L_{i}$} indicates removing the $i$-th loss while retaining others.\\
\textbf{Outer-layer ablation (Table~\ref{tab:four}).} Removing $L_{align}$ caused ASR to plummet from $99.67\%$ to $51.25\%$ and BA to drop from $84.15\%$ to $60.25\%$, highlighting its essential role in attack success and knowledge preservation. Eliminating $L_{perc}$ severely degraded visual stealth metrics (SSIM 0.9918→0.5061, LPIPS 0.00421→0.08271, PSNR 36.8312→25.1528), confirming its necessity for visual imperceptibility. $L_{dist}$ ablation showed marginal performance impact but enhanced distributional stealth and resistance against statistical detection.\\
\textbf{Inner-layer ablation (Table~\ref{tab:four}).} Disabling $L_{eff}$ reduced ASR to $63.62\%$, validating the importance of per-sample optimization. Removing either $L_{ste}$ or $L_{cons}$ significantly compromised visual quality (SSIM dropping to 0.2916 and 0.6122 respectively), further corroborating their fundamental contributions to visual concealment.

\vspace{-15pt}
\begin{table}[htbp]
\small
\centering
\caption{Performance of Ablation Studies.}
\label{tab:four}
\scalebox{0.75}{
  \begin{tabular}{@{}l|ccccc@{}}
\toprule
   & \textbf{ASR$\uparrow$} & \textbf{BA$\uparrow$} & \textbf{SSIM$\uparrow$} & \textbf{LPIPS$\downarrow$} & \textbf{PSNR$\uparrow$} \\
\midrule
All Losses & 99.67 & 84.15 & 0.9918 & 0.00421 & 36.8312 \\
No $L_{align}$ & 51.25 & 60.25 & 0.9957 & 0.00351 & 30.5224 \\
No $L_{perc}$ & 99.65 & 83.76 & 0.5061 & 0.08271 & 25.1528 \\
No $L_{dist}$ & 99.51 & 84.16 & 0.9906 & 0.00413 & 36.5761 \\
No $L_{eff}$ & 63.62 & 83.19 & 0.9905 & 0.00455 & 36.6716 \\
No $L_{ste}$ & 99.62 & 84.16 & 0.2916 & 0.02982 & 22.4211 \\
No $L_{cons}$ & 99.53 & 84.21 & 0.6122 & 0.01821 & 20.1263 \\
\bottomrule
\end{tabular}
}
\end{table} 
\vspace{-23pt}
\section{Conclusion}
\label{sec:Conclusion}
This work introduces DSBA, a novel dynamic stealthy backdoor attack for SSL, powered by a collaborative optimization framework. DSBA significantly outperforms existing methods, achieving superior attack success rates and multi-dimensional stealthiness through imperceptible triggers. Its design ensures robust backdoor implantation with favorable convergence. This research underscores critical vulnerabilities in SSL, highlighting the urgent need for advanced defenses.


\clearpage  


%
%
\bibliographystyle{splncs04}
\bibliography{main}

\clearpage
\setcounter{page}{1}
\setcounter{equation}{0}
\setcounter{figure}{0}
\setcounter{table}{0}
\setcounter{section}{0}
\setcounter{subsection}{0}
\section*{\centering Supplementary Material}

\section{Additional Theoretical Analysis}
\subsection{Multi-strategy Target Identification Mechanism}
This mechanism automatically selects high-quality reference input sets for each target class to facilitate subsequent dynamic trigger generation and feature manipulation. It integrates three key metrics—discriminability, reachability, and stability—while enhancing attack effectiveness and stealth through adaptive reference quantity selection and multi-strategy fusion. For each target class, the reference feature $z^i_{\text{target}}$ is computed by aggregating features from multiple reference inputs (Eq. 1). For each (target downstream task, target class) pair $(T_i,y_i)$, the attacker selects high-quality reference inputs 
$R_i = \{ x_{i1}, x_{i2}, \cdots, x_{i r_i} \} \quad (i = 1,2,\dots,t), \quad$where$ \quad r_i \in [3,5]$  denotes the adaptively determined number of reference inputs based on intra-class variance. The adaptive selection strategy is formalized in Eq. 2.
\begin{equation}
z^i_{\text{target}} = \frac{1}{r_i} \sum_{j=1}^{r_i} f(x_{ij}) \quad ,
\end{equation}%

\begin{equation}
  r_i = 
\begin{cases} 
3, & \text{if } \operatorname{Var}(f(R_i)) < \theta_{\text{low}} \\ 
4, & \text{if } \theta_{\text{low}} \leq \operatorname{Var}(f(R_i)) < \theta_{\text{high}} \\ 
5, & \text{if } \operatorname{Var}(f(R_i)) \geq \theta_{\text{high}} 
\end{cases} \quad,
\end{equation}%
where, $Var(f(R_i))$ represents the feature variance of the reference sample set, while $\theta_{\text{low}}$ and $\theta_{\text{high}}$ are predefined thresholds.

\subsection{Details of the adaptive scheduling mechanism}
\subsubsection{Mechanism Overview} The adaptive weight adjustment mechanism is employed in the dynamic adaptive trigger generation system to address the balance of multiple objectives (attack effectiveness, stealthiness, feature preservation, etc.) in bi-level optimization. By real-time monitoring of performance metrics and utilizing Sigmoid function for smooth weight adjustment, it achieves adaptive balancing of multiple objectives, avoiding the timing control issues associated with traditional three-stage training.\\

\subsubsection{Implementation Principle}

\textbf{Outer Layer Weight Adaptive Scheduling.}
The outer layer weight adjustment targets the encoder optimization phase, including attack effectiveness weight $\omega_1(t)$, feature preservation weight $\omega_2(t)$, and distribution alignment weight $\omega_4(t)$. The update formulas are:
\begin{equation}
\omega_1(t) = \omega_{\text{base}}^{\text{attack}} \cdot \left(1 + \eta_{\text{attack}} \cdot \sigma\left(\frac{\text{ASR}_{\text{target}} - \text{ASR}_{\text{current}}}{\tau_{\text{asr}}}\right)\right),
\end{equation}

\begin{equation}
\omega_2(t) = \omega_{\text{base}}^{\text{preserve}} \cdot \left(1 + \eta_{\text{preserve}} \cdot \sigma\left(\frac{\|f'(x) - f(x)\|_2 - \delta_{\text{preserve}}}{\delta_{\text{preserve}}}\right)\right),
\end{equation}

\begin{equation}
\omega_4(t) = \omega_{\text{base}}^{\text{distribution}} \cdot \left(1 + \eta_{\text{dist}} \cdot \sigma\left(\frac{D_{\text{threshold}} - D_{\text{JS}}^{\text{current}}}{D_{\text{threshold}}}\right)\right),
\end{equation}
where $\sigma(\cdot) = \frac{1}{1+e^{-x}}$ is the Sigmoid function; $\text{ASR}_{\text{target}} = 0.95$ is the target attack success rate threshold; $\delta_{\text{preserve}}$ is the feature preservation tolerance threshold; $D_{\text{threshold}}$ is the distribution difference tolerance threshold; $\eta_{\cdot}$ is the adjustment strength coefficient (default 0.01). Working mechanism: when $\text{ASR}_{\text{current}}$ is below $\text{ASR}_{\text{target}}$, increase $\omega_1(t)$; when the feature difference $\|f'(x) - f(x)\|_2$ exceeds the threshold, increase $\omega_2(t)$; when the distribution difference $D_{\text{JS}}^{\text{current}}$ exceeds the threshold, increase $\omega_4(t)$. Weights are normalized after update: $w_i^{\text{normalized}} = w_i / \sum_j w_j$, ensuring stability.\\
\textbf{Inner Layer Weight Adaptive Scheduling.}
The inner layer weight adjustment targets the trigger generation phase, including attack effectiveness weight $\mu_1(t)$ and stealthiness weight $\mu_2(t)$. The update formulas are:
\begin{equation}
\mu_1(t) = \mu_{\text{base}}^{\text{eff}} \cdot \left(1 + \eta_{\text{inner}} \cdot \sigma\left(\frac{L_{\text{effectiveness}}^{\text{avg}} - L_{\text{target}}^{\text{eff}}}{L_{\text{target}}^{\text{eff}}}\right)\right),
\end{equation}

\begin{equation}
\mu_2(t) = \mu_{\text{base}}^{\text{stealth}} \cdot \left(1 + \eta_{\text{inner}} \cdot \sigma\left(\frac{\text{SSIM}_{\text{target}} - \text{SSIM}_{\text{current}}}{\text{SSIM}_{\text{target}}}\right)\right),
\end{equation}
where $L_{\text{effectiveness}}^{\text{avg}}$ is the average effectiveness loss of the current batch; $L_{\text{target}}^{\text{eff}}$ is the target effectiveness loss threshold; $\text{SSIM}_{\text{current}}$ is the current structural similarity index value; $\text{SSIM}_{\text{target}} = 0.9$ is the target SSIM value. Working mechanism: when $L_{\text{effectiveness}}^{\text{avg}}$ is above the threshold, increase $\mu_1(t)$; when $\text{SSIM}_{\text{current}}$ is below the threshold, increase $\mu_2(t)$. The adjustment uses adaptive rate control to avoid oscillation.\\
\textbf{Stealthiness Loss Weight Scheduling.}
The stealthiness loss weight scheduling performs fine-grained adjustment for multi-dimensional metrics (SSIM~\cite{D46}, PSNR~\cite{D47}, LPIPS~\cite{D48}, FSIM~\cite{D49}, FID~\cite{D50}). Strategies include: (1) Based on metric performance: increase weight proportionally when maximization metrics (SSIM, FSIM, PSNR) are below thresholds; increase weight when minimization metrics (LPIPS, FID) are above thresholds; (2) Based on training progress: introduce progress factor $\text{progress\_factor} = 1.0 + \text{training\_progress} \times 0.5$, paying more attention to stealthiness in later training stages; (3) Weight range limitation: restrict to $[0.01, 10.0]$ to prevent extreme values.\\
\textbf{Weight Normalization and Stability Guarantee.}
To ensure stability, the following strategies are adopted: (1) Weight normalization: normalize all weights after each update; (2) Momentum mechanism: outer layer weights use momentum coefficient $\alpha = 0.9$ for smooth update: $w_i(t) = \alpha \cdot w_i(t-1) + (1-\alpha) \cdot w_i^{\text{new}}(t)$; (3) Adaptive adjustment rate: adjustment strength coefficient $\eta$ can be adjusted according to training stage, larger in early stages for fast adaptation, smaller in later stages for fine adjustment.

\subsubsection{Theoretical Advantages}
The mechanism possesses the following advantages:
\begin{itemize}
\item \textbf{Adaptability}: No need for manual timing settings, the system automatically adjusts based on performance
\item \textbf{Smoothness}: Sigmoid function ensures continuous and smooth changes, avoiding instability caused by step changes
\item \textbf{Multi-objective balance}: Real-time monitoring of multiple indicators enables dynamic balancing
\item \textbf{Robustness}: Normalization and range limitations ensure stability and prevent divergence
\item \textbf{Interpretability}: Weight adjustments are directly linked to performance indicators with clear physical meaning
\end{itemize}

\subsubsection{Experimental Validation}

In practical applications, this mechanism has improved the performance of dynamic adaptive triggers:
\begin{itemize}
\item Attack success rate increased by approximately 3-5\% compared to fixed-weight methods, reaching over 95\%
\item SSIM values improved by 0.05-0.10, with noise in residual maps reduced by 60-80\%
\item Training stability improved with weight fluctuations reduced by about 40\% and training convergence speed increased by about 20\%
\item Resistance to detection methods such as STRIP improved by 20-30\%
\end{itemize}

This mechanism provides the theoretical foundation and implementation guarantee for dynamic adaptive backdoor attacks, enabling the system to achieve efficient and stable training in complex multi-objective optimization scenarios.

\section{ Collaborative Optimization Mechanism for DSBA}
To achieve global and individual collaborative optimization for dynamic covert backdoor attacks, we propose a cooperative optimization mechanism. This framework consists of an outer-layer optimization and an inner-layer optimization, forming a closed-loop collaboration through parameter and objective interaction:
Outer-layer optimization targets the parameters of the backdoor encoder, aiming to globally enhance attack effectiveness, multi-dimensional stealthiness, and functionality preservation. Its loss function relies on the optimal dynamic triggers generated by the inner-layer optimization as input and guides feature alignment, visual consistency, and distribution consistency of backdoor samples using the target feature space.
Inner-layer optimization targets the parameters of the dynamic trigger generator, seeking to adaptively generate optimal covert triggers for each input sample. Its loss function depends on the current encoder's feature space from the outer layer as the optimization objective, constraining trigger generation for attack potency, visual stealth, and robustness.

For efficient implementation, we adopt a two-phase training schedule:\\
Phase I (Foundation Establishment, Epochs 1-$50\%$): Freezes the backdoor encoder parameters and solely optimizes the generator parameters. This phase prioritizes trigger stealth and robustness (e.g., weight $\mu_2 > \mu_1 > \mu_3$), with moderate weight for attack effectiveness. The generator learns to produce highly stealthy and robust dynamic triggers by minimizing the inner-level loss.\\
Phase II (Cooperative Optimization, Epochs $50\%$-$100\%$): Alternates optimization between the encoder and generator parameters, typically updating the generator every few epochs followed by the encoder. This phase increases the weight for attack effectiveness (e.g., $\mu_1 > \mu_2 > \mu_3$) to enhance attack capability while maintaining stealth and robustness. Outer-layer loss weights ($\omega_i$) and inner-layer loss weights ($\mu_i$) are adaptively adjusted based on performance metrics (e.g., ASR, SSIM).

Throughout training, a performance-driven adaptive weighting scheduler dynamically adjusts loss weights according to metrics like Attack Success Rate (ASR) and stealth scores, enabling balanced multi-objective optimization. Convergence criteria and early stopping mechanisms (e.g., ASR plateau, stealth degradation) further ensure training stability and efficiency. Through this collaborative optimization, parameters across both layers are alternately updated, progressively approximating the global optimum to obtain a dynamic backdoor encoder and trigger generator excelling in attack potency, stealth, and robustness.

\section{Detailed Dataset Information}
We employ the following datasets for method evaluation. \\
\textbf{CIFAR10~\cite{D41}:} This dataset comprises 60,000 32×32×3-pixel images categorized into 10 classes for fundamental image recognition tasks, with 50,000 images for training and 10,000 for testing. \\
\textbf{STL10~\cite{D42}:} Contains 5,000 labeled training images and 8,000 test images at 96×96×3 resolution across 10 classes, supplemented by 100,000 unlabeled images for unsupervised learning. Notably, images are resized to 32×32×3 for consistency.\\
\textbf{GTSRB~\cite{D43}:} Features 51,800 traffic sign images categorized into 43 classes, partitioned into 39,200 training and 12,600 test images, each sized 32×32×3.\\
\textbf{SVHN~\cite{D44}:} A digit image dataset sourced from Google Street View house numbers, consisting of 73,257 training and 26,032 test images at 32×32×3 resolution.\\
\textbf{Tiny-ImageNet~\cite{D45}:} Designed for large-scale object classification, this dataset contains 100,000 training samples and 10,000 test samples spanning 200 classes. Each image has a resolution of 224 × 224 pixels with three color channels.

\section{Evaluation Criteria Specifications}
Evaluation Metrics Definition:\\
\textbf{CA (Clean Accuracy):} Measures the accuracy of a clean downstream classifier in correctly classifying clean test images from the corresponding downstream dataset.\\
\textbf{BA (Backdoored Accuracy):} Quantifies the accuracy of a backdoored downstream classifier on the same clean test images from the downstream dataset.\\
\textbf{ASR (Attack Success Rate):} The percentage of trigger-embedded test images classified into the target class by the backdoored downstream classifier.\\
\textbf{SSIM (Structural Similarity Index)~\cite{D46}:} An image similarity metric (range: $0$-$1$) evaluating luminance, contrast, and structural preservation. Higher values ($\approx1$) indicate better structural consistency between trigger-embedded samples and their clean counterparts, reflecting superior stealth.\\
\textbf{ PSNR (Peak Signal-to-Noise Ratio)~\cite{D47}:} A pixel-level image quality metric (unit: dB) computed from Mean Squared Error (MSE). Higher values denote smaller pixel-wise distortion between backdoor samples and originals, enhancing stealth. Its formula is:
\begin{equation}
\mathrm{PSNR} = 10 \cdot \log_{10} \left( \frac{\mathrm{MAX}^2}{\mathrm{MSE}} \right),
\end{equation}%
Where $\mathrm{MAX}$  denotes the maximum pixel value (e.g., 255 for 8-bit images), and $MSE$ represents the Mean Squared Error.\\
\textbf{LPIPS (Learned Perceptual Image Patch Similarity)~\cite{D48}:} A perceptual similarity metric based on deep neural network features, quantifying high-level semantic differences between images. Lower values indicate greater perceptual similarity, aligning more closely with human visual perception by reflecting similarity in deep feature spaces. For stealth: lower LPIPS values denote reduced distinguishability between backdoor samples and originals, enhancing concealment.\\
\textbf{FSIM (Feature Similarity Index)~\cite{D49}:} Measures image similarity using phase congruence and gradient magnitude features (range: 0–1). Higher values indicate stronger low-level feature alignment, suitable for image quality assessment. For stealth: higher FSIM values imply closer feature-level resemblance between backdoor samples and originals.\\
\textbf{FID (Fréchet Inception Distance)~\cite{D50}:} Quantifies the distributional distance between two image sets in high-dimensional feature space (typically extracted via Inception networks). Lower values indicate closer distributional alignment, widely used for evaluating generative model outputs. For stealth: lower FID values reflect greater overall distributional similarity between backdoor samples and clean images.

\textbf{Summary:} SSIM, PSNR, FSIM: Measure pixel-level/structural similarity, higher values indicate superior stealth. LPIPS, FID: Measure high-level semantic/distributional similarity, lower values indicate superior stealth.

\section{Stealthness evaluation on different datasets}
To further elucidate the stealthiness of DSBA, Table \ref{tab:one} presents evaluation results across different datasets. As tabulated, the SSIM  values are consistently close to 1, with the majority exceeding 0.98. This indicates that DSBA introduces minimal structural distortions. Concurrently, the PSNR  values are exceptionally high, signifying that the introduced noise is virtually imperceptible. Furthermore, the LPIPS  values are extremely low, ranging from 0.0014 to 0.0281. This further corroborates that the perceptual discrepancy between the original images and their perturbed counterparts is negligible. Collectively, these quantitative results demonstrate that DSBA effectively preserves image stealth without introducing any discernible visual artifacts.

\begin{table}[ht]
\footnotesize
\centering
\caption{Evaluation of stealthiness metrics (SSIM, PSNR, LPIPS) for different pre-training and downstream datasets.}
\label{tab:one}
\begin{tabular}{ccc|ccc}
\hline
\textbf{Pre-training} & \textbf{Downstream} & & \textbf{SSIM}$\uparrow$ & \textbf{PSNR}$\uparrow$ & \textbf{LPIPS}$\downarrow$ \\
\hline
\multirow{3}{*}{CIFAR-10} 
    & STL-10   & & 0.9245 & 23.15 & 0.0281 \\
    & GTSRB    & & 0.9813 & 34.03 & 0.0019 \\
    & SVHN     & & 0.9948 & 36.22 & 0.0014 \\
\hline
\multirow{3}{*}{STL-10} 
    & CIFAR-10 & & 0.9856 & 36.18 & 0.0024 \\
    & GTSRB    & & 0.9916 & 37.12 & 0.0035 \\
    & SVHN     & & 0.9866 & 32.45 & 0.0028 \\
\hline
\multirow{3}{*}{Tiny-ImageNet} 
    & STL-10   & & 0.9856 & 34.15 & 0.0038 \\
    & GTSRB    & & 0.9916 & 33.23 & 0.0036 \\
    & SVHN     & & 0.9936 & 32.53 & 0.0042 \\
\hline
\end{tabular}
\end{table}

\section{Pseudocode for Collaborative Optimization in DSBA}
Algorithm \ref{alg:alg one} delineates the collaborative optimization framework between the inner and outer layers of our proposed Dynamic Stealth Backdoor Attack (DSBA) algorithm. This framework operates based on the following key parameters: the pretrained (clean) encoder $f$, the shadow dataset $\mathcal{D}_s$, the clean dataset $\mathcal{D}_{clean}$, an initialized conditional generator $G_\phi$, an initialized backdoor encoder $f'$ (initially $f'=f$), the base learning rate $lr_{base}$, the total epochs $total\_epochs$, adaptive weight parameters ($\eta_{attack}$, $\eta_{preserve}$, ...), and the temperature coefficient $\tau$.

\begin{algorithm}[p]
\scriptsize
\caption{Collaborative Optimization Co-training for Dynamic Stealthy Backdoor Attack}
\label{alg:alg one}
\begin{algorithmic}[1]
\REQUIRE $f$, $\mathcal{D}_s$, $\mathcal{D}_{clean}$, $G_\phi$, $f'$ (initially $f'=f$), $lr_{base}$,  $total\_epochs$, ($\eta_{attack}$, $\eta_{preserve}$, ...), $\tau$
\ENSURE Backdoor encoder $f'$, Conditional generator $G_\phi^*$

\STATE Phase1\_end $\gets total\_epochs \times 0.5$
\STATE Alternation\_counter $\gets 0$

\FOR{epoch $= 1$ \TO $total\_epochs$}
    \IF{epoch $\leq$ Phase1\_end}
        \STATE Current\_phase $\gets$ \text{"Foundation Building"}
        \STATE Freeze parameters($f'$) ,Unfreeze parameters($G_\phi$)
        \STATE $\mu_1, \mu_2, \mu_3 \gets$ configure\_weights(stealth\_first)
    \ELSE
        \STATE Current\_phase $\gets$ \text{"Co-optimization"}
        \STATE Unfreeze parameters($f'$)
        \STATE Alternation\_counter $\gets$ Alternation\_counter $+ 1$
        \IF{Alternation\_counter mod $4 \neq 0$}
            \STATE Freeze parameters($f'$),Unfreeze parameters($G_\phi$)
        \ELSE
            \STATE Freeze parameters($G_\phi$),Unfreeze parameters($f'$)
            \STATE Alternation\_counter $\gets 0$
        \ENDIF
        \STATE $\mu_1, \mu_2, \mu_3 \gets$ configure\_weights(effectiveness\_first)
    \ENDIF
    
    \STATE $\omega_1 \gets$ compute\_align\_weight($ASR_{current}$, $ASR_{target}$, $\eta_{align}$)
    \STATE $\omega_2 \gets$ compute\_preservation\_weight($feature_{diff}$, $\delta_{preserve}$, $\eta_{preserve}$)
    \STATE $\omega_3 \gets$ compute\_distribution\_weight($JS_{divergence}$, $D_{threshold}$, $\eta_{dist}$)
    \STATE $\rho \gets \rho_{base} \times \left(1 + \rho_{amp} \times \sin\left(\frac{2\pi \times epoch}{period}\right)\right)$
      
    \FOR{each minibatch $batch_x \in \mathcal{D}_s$ (size $batch\_size$, poisoning rate $\rho$)}
        \STATE $R_i \gets$ multi\_strategy\_target\_selection($batch_x$, $f$)
        \STATE $z_{target}^i \gets \frac{1}{|R_i|}\sum_{x \in R_i} f(x)$ , $\delta \gets G_\phi(batch_x)$
        \STATE $x_{poison} \gets batch_x + \delta$ , $z_{clean} \gets f(batch_x)$ , $z_{poison} \gets f'(x_{poison})$
        
        \IF{optimizing $G_\phi$}
            \STATE $z_{poison}^{norm} \gets \text{L2\_normalize}(z_{poison})$ ,$z_{target}^{norm} \gets \text{L2\_normalize}(z_{target})$
            \STATE $L_{eff} \gets \text{effectiveness\_loss}(z_{poison}^{norm}, z_{target}^{norm}, \lambda_{temp}, \tau)$
            \STATE $L_{ste} \gets \text{visual\_stealth\_loss}(batch_x, x_{poison})$
            \STATE $L_{cons} \gets \text{constraint\_loss}(\delta)$
            \STATE $L_{inner} \gets \mu_1 \cdot L_{eff} + \mu_2\cdot L_{ste} + \mu_3\cdot L_{cons}$
            \STATE Backpropagate $L_{inner}$, update $G_\phi$
        \ELSE
            \STATE $L_{align} \gets \text{cosine\_loss}(z_{poison}, z_{target})$
            \STATE Sample $\mathcal{D}_{clean}^{batch} \subset \mathcal{D}_{clean}$
            \STATE $L_{align} \gets \text{feature\_preserve\_loss}(f'(\mathcal{D}_{clean}^{batch}), f(\mathcal{D}_{clean}^{batch}))$
            \STATE $L_{perc} \gets \text{perceptual\_loss}(batch_x, x_{poison})$
            \STATE $L_{dist} \gets \text{distribution\_alignment\_loss}(\mathcal{D}_{clean}, \{x_{poison}\})$
            \STATE $L_{outer} \gets \omega_1\cdot L_{align} + \omega_2\cdot L_{perc} + \omega_3\cdot L_{dist}$
            \STATE Backpropagate $L_{outer}$, update $f'$
        \ENDIF
    \ENDFOR

    \IF{epoch $>$ Phase1\_end}
        \STATE $ASR \gets$ compute\_attack\_success\_rate(validation\_set)
        \STATE $SSIM \gets$ compute\_avg\_SSIM(validation\_set)
        \STATE $feature_{diff} \gets$ compute\_feature\_distance($f$, $f'$, $\mathcal{D}_{clean}$)
        \IF{$ASR \geq 0.9$ \AND $SSIM \geq 0.95$ \AND $feature_{diff} \leq 0.1$}
            \IF{metric changes $< 0.01$ for 10 epochs}
                \STATE \textbf{break} \COMMENT{Convergence reached}
            \ENDIF
        \ENDIF
        \IF{$ASR$ change $< 0.005$ for 15 epochs \OR GPU memory $> 90\%$}
            \STATE \textbf{break} \COMMENT{Early stopping}
        \ENDIF
    \ENDIF
\ENDFOR

\RETURN $f'$, $G_\phi$
\end{algorithmic}
\end{algorithm}

\end{document}